\documentclass[conference]{IEEEtran}
\IEEEoverridecommandlockouts
\usepackage{cite}
\usepackage{amsmath,amssymb,amsfonts}
\usepackage{algorithmic}
\usepackage{graphicx}
\usepackage{textcomp}
\usepackage{xcolor}
\def\BibTeX{{\rm B\kern-.05em{\sc i\kern-.025em b}\kern-.08em
  T\kern-.1667em\lower.7ex\hbox{E}\kern-.125emX}}
\begin{document}

\title{Risk-Neutral Pricing Model of Uniswap Liquidity Providing Position: A Stopping Time Approach\\}

\author{\IEEEauthorblockN{HOU Liang}
\IEEEauthorblockA{\textit{Antalpha labs} \\
\textit{Antalpha}\\
Beijing, China \\
liang.hou@antalpha.com}
\and
\IEEEauthorblockN{YU Hao}
\IEEEauthorblockA{\textit{Antalpha labs} \\
\textit{Antalpha}\\
Hangzhou, China \\
yuhao0126liuer@gmail.com}
\and
\IEEEauthorblockN{XU Guosong}
\IEEEauthorblockA{\textit{Derivatives Trading Division} \\
\textit{China Merchants Securities Co., Ltd.}\\
Shenzhen, China \\
xuguosong@cmschina.com.cn}
}

\maketitle

\begin{abstract}
In this paper, we introduce a novel pricing model for Uniswap V3, built upon stochastic processes and the Martingale Stopping Theorem. This model innovatively frames the valuation of positions within Uniswap V3. We further conduct a numerical analysis and examine the sensitivities through Greek risk measures to elucidate the model's implications. Additionally, we demonstrate the model's practical application by construct the implied volatility of the model and compare it to other exist model. The results underscore the model’s significant academic contribution and its practical applicability for Uniswap liquidity providers (LPs), particularly in assessing risk exposure and guiding hedging strategies.
\end{abstract}


\section{Introduction}
\subsection{Background and Motivation}
Uniswap, launched in 2018, is one of the leading decentralized exchanges (DEXs) in the cryptocurrency market, enabling users to swap tokens directly from their wallets through liquidity pools, without relying on traditional intermediaries. As the first automated market maker (AMM)\cite{b1} to achieve widespread adoption, Uniswap has played a pivotal role in the growth of decentralized finance (DeFi).

The introduction of Uniswap V3 marked a significant advancement in decentralized exchange technology, most notably through the concept of concentrated liquidity. This feature allows LPs to allocate capital to specific price ranges, improving capital efficiency and enhancing potential returns while simultaneously reducing price risk.

With the ongoing growth of DeFi adoption, LPs now face a more complex landscape, balancing a range of factors such as risk, returns, and volatility. This requires a sophisticated understanding of liquidity provision strategies.

In this work, we treat Uniswap V3 as an exotic perpetual derivative and use the Black-Scholes framework, along with the Martingale Stopping Theorem, to derive a closed-form pricing formula. Additionally, we perform a Monte Carlo simulation and sensitivity analysis to evaluate the model’s performance. We also compare our model with other state-of-the-art models to assess its advantages and limitations.

Looking ahead, Uniswap V4 is expected to introduce even more advanced features that will further enhance the efficiency and flexibility of decentralized exchange mechanisms.  Our model is designed to incorporate these new features of Uniswap V4, ensuring its adaptability to future developments.

\subsection{Related Work}

Our work builds on and extends two broad strands of literature: (i) DeFi research on AMMs, and (ii) traditional financial engineering studies on market-making. Below, we discuss the most relevant works and highlight how our research contributes to these fields.

\subsubsection{Automated Market Makers}

A significant advancement in AMM research is the concept of Loss-Versus-Rebalancing (LVR), introduced by \cite{b2} to quantify the adverse selection costs faced by LPs in AMM. LVR measures the performance gap between an LP's portfolio and a dynamically rebalanced portfolio, capturing the losses incurred due to stale prices exploited by arbitrageurs.

Another important contribution is the work of \cite{b3}, who modeled Uniswap V3 liquidity positions as perpetual options within the Black-Scholes-Merton (BSM) framework. Lambert's work treats LP positions as covered calls and explores implied volatility and hedging strategies.

Recent studies have further advanced the understanding of AMMs. For example, \cite{b4} analyze the tradeoff between risks and returns for LPs in Uniswap V3, while \cite{b5} study the predictable losses of LPs in a continuous-time. \cite{b6} explore the hedging of impermanent losses. Closest to our work are the models in \cite{b7} and \cite{b8}, which focus on fee revenue and use approximation techniques to obtain dynamic strategies.

\subsubsection{Market-Making in Traditional Finance}

Our work is also related to the market-making literature in traditional finance. Early works include \cite{b9} and \cite{b10}, which model market-making as an inventory management problem and derive optimal bid-ask spread strategies. These foundational studies have been extended in many directions; see \cite{b11} and \cite{b12} for recent developments

Another particularly relevant area is the study of perpetual derivatives, which share similarities with the perpetual nature of Uniswap V3 liquidity positions. \cite{b13} provides a comprehensive treatment of perpetual put options using stochastic processes, deriving the distribution of stopping times and optimal stopping strategies.

\subsection{Contributions}
Our work is structured into four key steps, each addressing a critical aspect of pricing Uniswap V3 liquidity positions and extending the model's applicability:

Step 1: Perpetual Liquidity Position Modeling

We model Uniswap V3 liquidity positions as perpetual derivatives, using the Martingale Stopping Theorem and Laplace Transform to derive closed-form pricing formulas. This approach captures the perpetual nature of liquidity positions and eliminates the need for fixed expiration times.

Step 2: Monte Carlo Validation

We validate our closed-form formulas through Monte Carlo simulations. The results align closely with the analytical solutions, confirming the accuracy and robustness of our model.

Step 3: Greeks and Implied Volatility

We conduct a comprehensive analysis of Greek risk measures (Delta, Gamma, Vega) and derive implied volatility from our model. A key comparison with the LVR framework highlights the advantages of our approach in capturing implied volatility.

Step 4: Generalization to Other AMMs and Uniswap V4

We discuss the potential extension of our model to other AMMs and its adaptation to the dynamic fee structures expected in Uniswap V4. While our current focus is on Uniswap V3, we outline possible modifications to the framework to accommodate future developments in AMM design.

\section{Derivation}

\subsection{Summary of Model Derivation}


Uniswap V3 can be divided into two primary pricing components: the LPs segment and the Rebate segment. The LPs component is structured to accommodate either European or American option styles. Under the European construct, the Uniswap contract realizes its terminal value solely upon the price breaching the predetermined upper or lower thresholds. In contrast, the American framework allows the contract holder the flexibility to terminate the automated market-making process at any chosen point in time.
The valuation of the Rebate segment hinges on the method selected by the contract holder for Rebate withdrawal. The upper limit scenario assumes the possibility of continuous Rebate extraction at the precise point of accrual. Meanwhile, the lower limit approach considers a singular, comprehensive withdrawal of all Rebates coinciding with the termination of the contract.

\subsection{Divide and Conquer}\label{AA}
We initially assume that the ETHUSDT price follows a Geometric Brownian Motion (GBM) characterized by drift $\mu$ and annualized volatility $\sigma$ :
$$dS(t)=\mu S(t)dt+ \sigma S(t)dB_t$$
or alternatively:
$$
S(t) = S_0\exp((\mu-\frac{\sigma^2}{2})t+\sigma B_t)
$$
Define the unit price of ETHUSDT as follows
$$P(t) = \frac{S(t)}{S_0}=\exp((\mu-\frac{\sigma^2}{2})t+\sigma B_t)$$
We partition the pricing structure of Uniswap V3 into two segments:
$$
V = V_{\text{LP}}+V_{\text{fee}}
$$
\subsubsection{Liquidity Provider segment}
The unit price of the LPs segment within the Uniswap V3 Contract, constrained by $S_H>S_0> S_L$, can be expressed as:
$$
\begin{aligned}
V_\text{LP}(P_t)&=
\begin{cases}
L_q P(t)(\frac{1}{\sqrt{L}}-\frac{1}{\sqrt{H}}), P(t)<L \\
L_q(2\sqrt{P(t)}-\sqrt{L}-\frac{P(t)}{\sqrt{H}}), L<P(t)<H \\
L_q(\sqrt{H}-\sqrt{L}), H<P(t)
\end{cases} \\
\end{aligned}
$$
where $H, L$ represent the unit bounds:
$$
\begin{aligned}
H &= \frac{S_H}{S_0}\\
L &= \frac{S_L}{S_0}\\
\end{aligned}
$$
Additionally, the Liquidity Parameter is defined as:
$$
L_q =\frac{1}{2-\sqrt{L}-\frac{1}{\sqrt{H}}}
$$
\subsubsection{Rebate segment}
The formulation for the rebate fee segment is comparatively straightforward:
$$
V_{\text{fee}} = C_a L_q V_{LP}(P_t) t=365\times C_d L_q V_{LP}(P_t) t
$$
where $C_a, C_d$ are the daily and annual rebate rate.
\subsection{European LP Pricing}\label{AA}
We are interested in evaluating:
$$
V(0) = \mathbb E(V(t)) = \mathbb E(V_{\text{LP}}(t))+\mathbb E(V_{\text{fee}}(t))
$$
For the Uniswap V3 opsition, the stopping time $\tau$ is defined as:
$$\tau = \inf\limits_{t<\infty}\{S(t)\leq S_L \text{ or }S(t)\geq S_H\}$$
If the contract cannot be terminated before the stopping time, the current valuation of Uniswap V3 can be formulated as:
$$
\begin{aligned}
V_{E}(0) &= \mathbb E(V(t))\\
&= \mathbb E(V_{\text{LP}}(t))+\mathbb E(V_{\text{fee}}(t))\\
&= \mathbb E[e^{-r\tau}V_\text{LP}(H)|S(\tau)=S_H]+ \mathbb E[e^{-r\tau}V_\text{LP}(L)|S(\tau)=S_L] \\& + \mathbb E(V_{\text{fee}}(t))\\
&= V_\text{LP}(H) \mathbb E[e^{-r\tau}|S(\tau)=S_H]+V_\text{LP}(L) \mathbb E[e^{-r\tau}|S(\tau)=S_L] \\
&+\mathbb E(V_{\text{fee}}(t))
\end{aligned}
$$
The normalized log-unit price process of ETHUSDT is given by:
$$
\bar W(t) = \frac{\ln P(t)}{\sigma}=(\frac{\mu}{\sigma}-\frac{\sigma}{2})t + B_t
$$
Let us denote:
$$
\mu' = \frac{\mu}{\sigma}-\frac{\sigma}{2}
$$
Thus, we have:
$$
\bar W(t) =x +\mu't +B_t
$$
where $x$ is the start point, $a,b$ are designated as the unit bounds:
$$
\begin{aligned}
b &= \frac{\ln H}{\sigma}, b' = b-x\\
a &= \frac{\ln L}{\sigma}, a' = x-a\\
\end{aligned}
$$
Moreover, it also holds that:
$$\tau = \inf\limits_{t<\infty}\{\bar W(t)\leq a \text{ or }\bar W(t)\geq b\}$$
Utilizing the Laplace Transform, we can express the following expectations:
$$
\mathbb E[e^{-r\tau}|S(\tau)=S_H] = \exp({\mu'b'})\frac{\sinh[a'\sqrt{\mu'^2+2r}]}{\sinh[(b-a)\sqrt{\mu'^2+2r}]}
$$
$$
\mathbb E[e^{-r\tau}|S(\tau)=S_L]=\exp({-\mu'a'})\frac{\sinh[(b'\sqrt{\mu'^2+2r}]}{\sinh[(b-a)\sqrt{\mu'^2+2r}]}
$$
\subsection{American LP Pricing}
In Euro-like pricing, liquidity providers can only exit when the price reaches the boundary. We further propose an American-like pricing model that allows liquidity providers to exit at any time.
As Uniswap V3 contracts are perpetual, the termination of the contract relies exclusively on price movements, independent of time $t$. Therefore, it can be assumed that there exist boundaries determined solely by price levels. This implies that exit strategies should be based only on price considerations, analogous to the derivation of the perpetual American put option in \cite{b13}'s work. So we have:
$$
L<L_1<x<L_2<H
$$
where terminating automated market-making at prices 
$L_1$ or $L_2$. A set of stopping boundaries $L_1, L_2$ can be determined through numerical optimization.
First, we denote $c,d$ as previous section:
$$
\begin{aligned}
d &= \frac{\ln L_2}{\sigma}, d' = d-x\\
c &= \frac{\ln L_1}{\sigma}, c' = x-c\\
\end{aligned}
$$
Then the equivalent price of the European contract:
$$
\begin{aligned}
V_E &= LP_{L_2} \exp({\mu'd'})\frac{\sinh[c'\sqrt{\mu'^2+2r}]}{\sinh[(d-c)\sqrt{\mu'^2+2r}]}\\
&+LP_{L_1}\exp({-\mu'c'})\frac{\sinh[(d'\sqrt{\mu'^2+2r}]}{\sinh[(d-c)\sqrt{\mu'^2+2r}]}    
\end{aligned} 
$$
The American contract can be priced using optimization techniques as following:

$$
V_A = \mathop{\max_{L<L_1<x<L_2<H}V_E(L_1,L_2)}
$$

\subsection{Rebate Pricing}
For the rebate segment, if we assume that we can withdraw rebates continuously,(it is an upper bound of rebate)
$$
\begin{aligned}
\mathbb E(V_{\text{fee}}(t)) & =\mathbb{E}\left[\int_0^\tau C_aL_q \cdot e^{-r t} d t\right] \\
& =C_aL_q \cdot \mathbb{E}\left[\int_0^\tau e^{-r t} d t\right] \\
& =C_aL_q \cdot \mathbb{E}\left[\frac{1-e^{-r \tau}}{r}\right] \\
& =\frac{C_aL_q}{r} \cdot (1-\mathbb{E}\left[{e^{-r \tau}}\right])
\end{aligned}
$$
where,
$$
\begin{aligned}
&\mathbb{E} [e^{-r \tau}]:=F(r)\\
&=\frac{e^{-\mu'a'} \operatorname{sh}\left(b' \sqrt{2 r+\mu'^2}\right)+e^{\mu' b'} \operatorname{sh}\left(a' \sqrt{2 r+\mu'^2}\right)}{\operatorname{sh}\left((b-a) \sqrt{2 r+\mu'^2}\right)}
\end{aligned}
$$
For the lower bound, we assume that we can withdraw all rebates when LPs close their position,
$$
\mathbb E(V_{\text{fee}}(t)) = \mathbb E(C_aL_qe^{-r\tau}\tau) = C_aL_q\mathbb E(e^{-r\tau}\tau)
$$
where,
$$
\begin{aligned}
&\mathbb E[e^{-r\tau}\tau]= \int_{\mathbb R}\tau e^{-r\tau}p(\tau)d\tau\\
 &=\int_{0}^{\infty}\tau p(\tau)e^{-r\tau}d\tau\\
 &= \mathcal L\{\tau p(\tau)\}(r)\\
 &= -\frac{d F(r)}{dr}\\
&=\frac{(b-a)(e^{-a'\mu'}\operatorname{sh}[b'\sqrt{2r+\mu'^2}]+e^{b'\mu}\operatorname{sh}[a'\sqrt{2r+\mu'^2}])}{\operatorname{th}[(b-a)\sqrt{2r+\mu'^2}]\operatorname{sh}[(b-a)\sqrt{2r+\mu'^2}](\sqrt{2r+\mu'^2}) }
 \\& \ \ \ \ -
\frac{b' e^{-a'\mu'}\operatorname{ch}[b'\sqrt{2r+\mu'^2}]+ae^{b\mu}\operatorname{ch}[a'\sqrt{2r+\mu'^2}]}{\operatorname{sh}[(b-a)\sqrt{2r+\mu'^2}](\sqrt{2r+\mu'^2})}
\end{aligned}
$$

\begin{figure}
    \centering
    \includegraphics[width=1\linewidth]{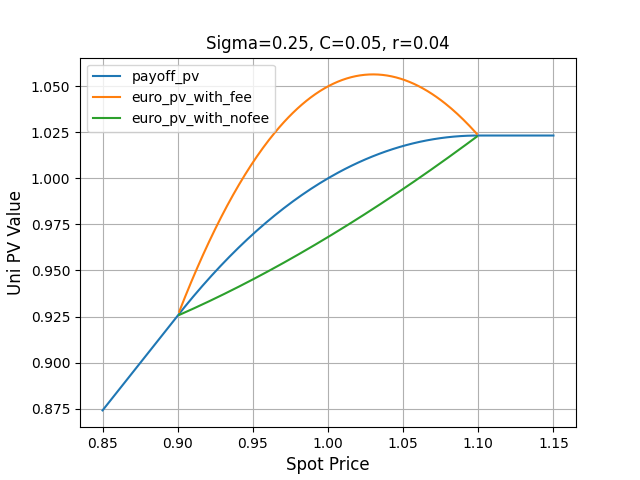}
    \caption{relation between spot price and present value}
    \label{fig:spot_pv}
\end{figure}

\begin{figure}
    \centering
    \includegraphics[width=1\linewidth]{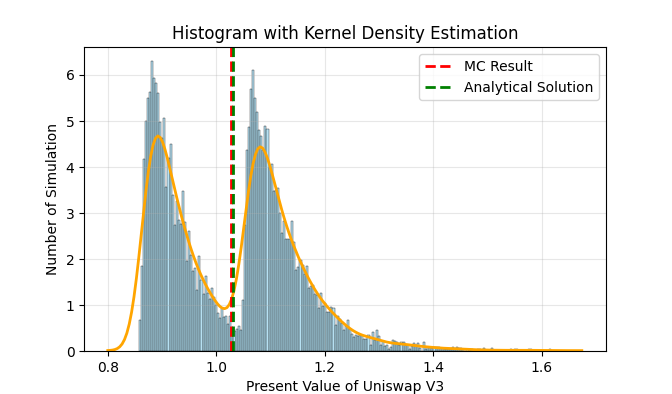}
    \caption{Euro-Uniswap price distribution: Monte Carlo vs analytical solution}
    \label{fig:hist}
\end{figure}

As illustrated in Figure~\ref{fig:spot_pv}, the blue line represents the variation of the payoff pricing model in response to changes in the spot price. The orange and green lines depict the corresponding numerical changes in the pricing model of a European option, where the orange line incorporates transaction costs and discounting, while the green line excludes these factors, both following the fluctuations in the spot price.We can observe that, under the current market parameters, if the option pricing model does not include transaction costs, it results in an expected loss at the current price, meaning that the present value is less than 1.Only by incorporating the potential transaction costs and their discounted value can we generate a positive return.

Figure~\ref{fig:hist} shows a Monte Carlo simulation of asset price dynamics under GBM, with $S_0=1, \sigma=0.6,\mu=0, r=0.04,H=1.2, L=0.8$.Running 10,000 simulations, we calculated payoffs based on boundary hits, discounted them, and matched the results to our analytical solution, confirming accuracy.

The outcome histogram exhibits a bimodal shape, reflecting price stops at $H$ or $L$ due to Uniswap V3's stopping conditions. The peaks correspond to boundary interactions, capturing the risk/reward trade-offs for liquidity providers in decentralized exchanges.

\subsection{Formulation of Case-Specific Models}
Based on the derivations from the foundational section, we have obtained the pricing formulas for the following four scenarios. For the pricing formula of the European LPs option problem under a GBM process, assuming that transaction fees can only be collected upon exiting the position, the pricing formula is as follows:
$$
\begin{aligned}
&V_t \\
&=LP_He^{\mu'b'}\frac{sh(a'\sqrt{2r+\mu'^2})}{sh((b-a)\sqrt{2r+\mu'^2})} \\
&+LP_Le^{\mu'a'}\frac{sh(b'\sqrt{2r+\mu'^2})}{sh((b-a)\sqrt{2r+\mu'^2})} \\
&+C_aL_q* \\
&(\frac{(b-a)(e^{\mu'a'}sh(b'\sqrt{2r+\mu'^2})+e^{\mu'b'}sh(a'\sqrt{2r+\mu'^2}))}{\sqrt{2r+\mu'^2}th((b-a)\sqrt{2r+\mu'^2})sh((b-a)\sqrt{2r+\mu'^2})} \\
&-\frac{e^{\mu'a'}b'ch(b'\sqrt{2r+\mu'^2})+e^{\mu'b'}a'ch(a'\sqrt{2r+\mu'^2})}{\sqrt{2r+\mu'^2}sh((b-a)\sqrt{2r+\mu'^2})})
\end{aligned}
$$

When we assume that transaction fees can be continuously collected during the lifetime of the position, the pricing formula becomes:
$$
\begin{aligned}
V_t
&=LP_He^{\mu'b'}\frac{sh(a'\sqrt{2r+\mu'^2})}{sh((b-a)\sqrt{2r+\mu'^2})} \\
&+LP_Le^{\mu'a'}\frac{sh(b'\sqrt{2r+\mu'^2})}{sh((b-a)\sqrt{2r+\mu'^2})} \\
&+\frac{C_aL_q}{r}(1 \\
&-\frac{e^{\mu'a'}sh(b'\sqrt{2r+\mu'^2})+e^{\mu'b'}sh(a'\sqrt{2r+\mu'^2})}{sh((b-a)\sqrt{2r+\mu'^2})})
\end{aligned}
$$
For the pricing formula of the American LPs option problem, assuming that transaction fees can only be collected upon exiting the position, the pricing formula is as follows:
$$
\begin{aligned}
&V_t=LP_{L_2}e^{\mu'd'}\frac{sh(c'\sqrt{2r+\mu'^2})}{sh((d-c)\sqrt{2r+\mu'^2})} \\
&+LP_{L_1}e^{\mu'c'}\frac{sh(bd'\sqrt{2r+\mu'^2})}{sh((d-c)\sqrt{2r+\mu'^2})} \\
&+C_aL_q(\frac{(d-c)(e^{\mu'c'}sh(d'\sqrt{2r+\mu'^2})+e^{\mu'd'}sh(c'\sqrt{2r+\mu'^2}))}{\sqrt{2r+\mu'^2}th((d-c)\sqrt{2r+\mu'^2})sh((d-c)\sqrt{2r+\mu'^2})} \\
&-\frac{e^{\mu'c'}d'ch(d'\sqrt{2r+\mu'^2})+e^{\mu'd'}c'ch(c'\sqrt{2r+\mu'^2})}{\sqrt{2r+\mu'^2}sh((d-c)\sqrt{2r+\mu'^2})})
\end{aligned}
$$
When we assume that transaction fees can be continuously collected during the lifetime of the position, the pricing formula for the American LPs option problem becomes:
$$
\begin{aligned}
V_t
&=LP_{L_2}e^{\mu'd'}\frac{sh(c'\sqrt{2r+\mu'^2})}{sh((d-c)\sqrt{2r+\mu'^2})} \\
&+LP_{L_1}e^{\mu'c'}\frac{sh(bd'\sqrt{2r+\mu'^2})}{sh((d-c)\sqrt{2r+\mu'^2})} \\
&+\frac{C_aL_q}{r}(1 \\
&-\frac{e^{\mu'c'}sh(d'\sqrt{2r+\mu'^2})+e^{\mu'd'}sh(c'\sqrt{2r+\mu'^2})}{sh((d-c)\sqrt{2r+\mu'^2})})
\end{aligned}
$$

\section{Sensitivity Analysis}
In the previous sections, we derived the analytical solution of Uniswap V3 positions. In this section, we focus on the sensitivity analysis of the Uniswap V3 position, specifically examining the Greeks associated with a typical European option framework. We previously derived the closed-form solutions for the pricing mechanisms of Uniswap V3 positions. Here, we extend our analysis to assess how variations in underlying parameters affect the pricing—commonly referred to as "Greeks."

\begin{table}[htbp] \caption{$C= 0.2$   $r= 0.05$  $\sigma= 0.7$} \centering \begin{tabular}{|c|c|c|c|c|c|} \hline \textbf{Model} & \textbf{PV} & \textbf{Delta} & \textbf{Gamma} & \textbf{Vega} & \textbf{Rho} \\ \hline Payoff & 1.000 & 0.561 & -0.632 & nan & nan \\ European & 1.422 & 0.427 & -0.747 & 1.720 & -25.892 \\ American & 1.696 & 0.539 & -0.989 & 0.843 & -19.425 \\ \hline
\end{tabular} 
\label{good market condition}
\end{table}

\begin{table}[htbp] \caption{$C= 0.04$  $r= 0.05$  $\sigma= 0.4$} \centering \begin{tabular}{|c|c|c|c|c|c|} \hline \textbf{Model} & \textbf{PV} & \textbf{Delta} & \textbf{Gamma} & \textbf{Vega} & \textbf{Rho} \\ \hline Payoff & 1.000 & 0.784 & -0.883 & nan & nan \\ European & 0.643 & 0.074 & 0.073 & 1.296 & -7.546 \\ American & 1.018 & 0.765 & -1.059 & 0.200 & -2.019 \\ \hline \end{tabular} 
\label{bad market condition}
\end{table}

\begin{figure}
    \centering
    \includegraphics[width=1\linewidth]{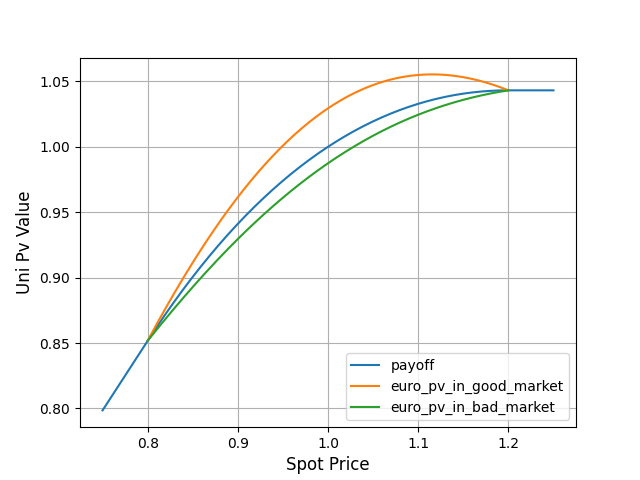}
    \caption{European LP Pricing Analysis}
    \label{fig:European LP Pricing Analysis}
\end{figure}

Our Greeks analysis will use two market parameter sets: one with high fees and high volatility ("good market"), and another with low fees and low volatility ("bad market"). Using the European LP option pricing model, we compare valuation differences between good/bad market conditions and payoff models across price levels, with a range order interval of (0.8, 1.2).

As shown in Figure 3, the European LP model reveals a significant gap between Uniswap V3's current payoff and its present value, as the latter includes discounted future fees. In low volatility and fee scenarios, not withdrawing the position leads to value depreciation, with future fees failing to offset losses (green line). Conversely, in high volatility and fee "good market" conditions, the position's actual value exceeds its payoff (orange line).

\begin{figure}
    \centering
    \includegraphics[width=1\linewidth]{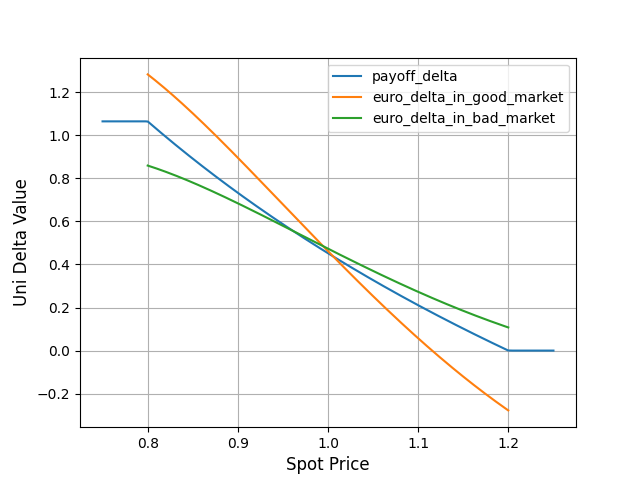}
    \caption{European LP Delta Analysis}
    \label{fig:European LP Delta Analysis}
\end{figure}

\subsection{Delta Analysis}
In Uniswap V3 trading, delta hedging is often used to manage delta-neutral risk. However, our European LP option pricing model reveals that delta risk within $(L, H)$ the range differs from payoff-based delta.

For example, in Figure~\ref{fig:European LP Delta Analysis}, the orange line (model delta) exceeds the blue line (payoff-based delta) at a price of 0.9, indicating higher actual delta risk and requiring more hedging. Conversely, at 1.1, model delta risk is lower, making excessive hedging unnecessary.

\begin{figure}
    \centering
    \includegraphics[width=1\linewidth]{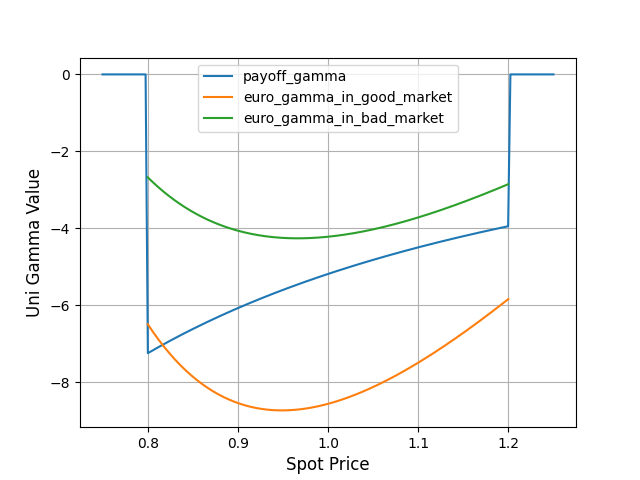}
    \caption{European LP Gamma Analysis}
    \label{fig:European LP Gamma Analysis}
\end{figure}

\subsection{Gamma Analysis}

Gamma analysis of the European LP pricing model versus the payoff-based model (Figure~\ref{fig:European LP Gamma Analysis}) shows that Uniswap V3 positions inherently carry gamma risk. Large gamma values cause delta to fluctuate significantly with small changes in 
$S$, reducing delta hedging effectiveness and requiring frequent adjustments. Conversely, small gamma values stabilize delta, minimizing hedging needs.

Figure~\ref{fig:European LP Gamma Analysis} reveals a significant gap between gamma risks from payoff-based and European LP models, highlighting the need for more refined risk control in Uniswap V3 positions.

\subsection{Vega Analysis}
\begin{figure}
    \centering
    \includegraphics[width=1\linewidth]{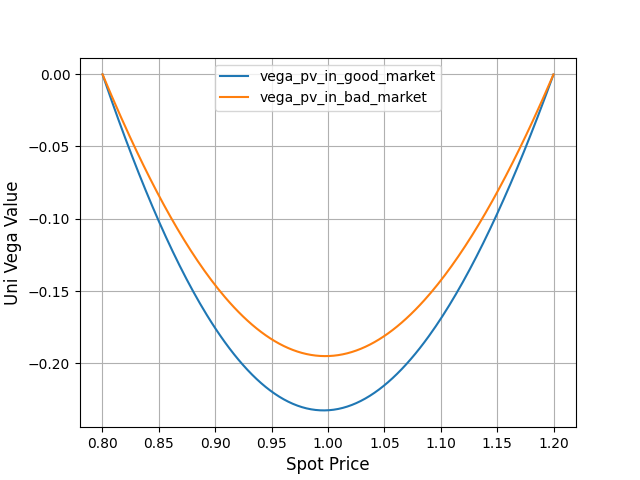}
    \caption{European LP Vega Analysis}
    \label{fig:European LP Vega Analysis}
\end{figure}

Vega measures the sensitivity of the valuation of a Uniswap V3 European pricing position to changes in volatility. From the Figure~\ref{fig:European LP Vega Analysis}, we observe the variation of vega with respect to spot prices under different market conditions. Notably, there is no vega estimation graph for the payoff-based pricing model, as this model assumes no correlation between price and volatility, and thus does not account for vega risk.

From the Figure~\ref{fig:European LP Vega Analysis}, it is evident that vega remains consistently negative within the defined $(L,H)$ range. This indicates that holding a Uniswap V3 position inherently represents a short volatility strategy, implying that lower volatility leads to higher returns.

\subsection{Greeks Comparison between different range order}
After analyzing Greeks under Uniswap V3 positions, we compare Greeks across different $(L,H)$ ranges to explore pricing relationships. We test ranges $(0.8, 1.1)$, $(0.8, 1.2)$, and $(0.8, 1.3)$ under$r=0.05$, $C=0.2$, $\mu=0$, and $\sigma=0.7$.

\begin{figure}
    \centering
    \includegraphics[width=1\linewidth]{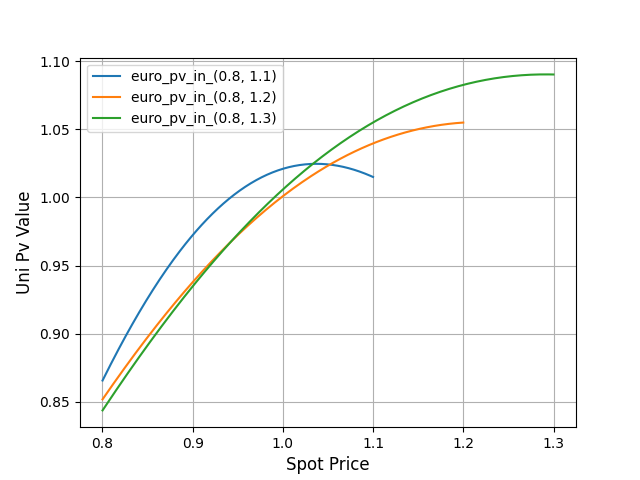}
    \caption{Different Range LP Pricing Comparison}
    \label{fig:Different Range LP Pricing Comparison}
\end{figure}

\begin{figure}
    \centering
    \includegraphics[width=1\linewidth]{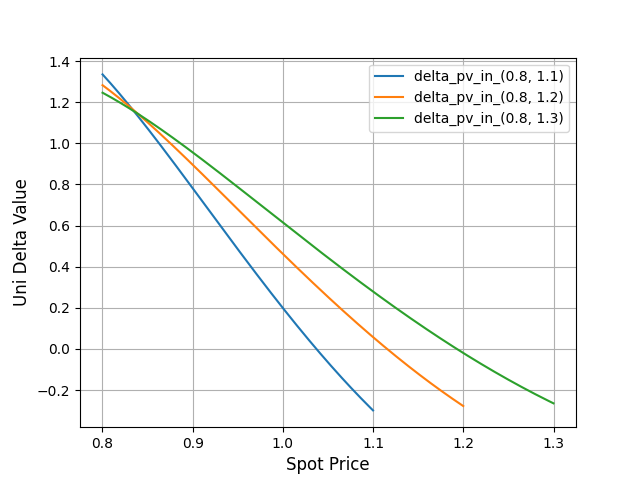}
    \caption{Different Range LP Pricing delta Comparison}
    \label{fig:Different Range LP Pricing delta Comparison}
\end{figure}

\begin{figure}
    \centering
    \includegraphics[width=1\linewidth]{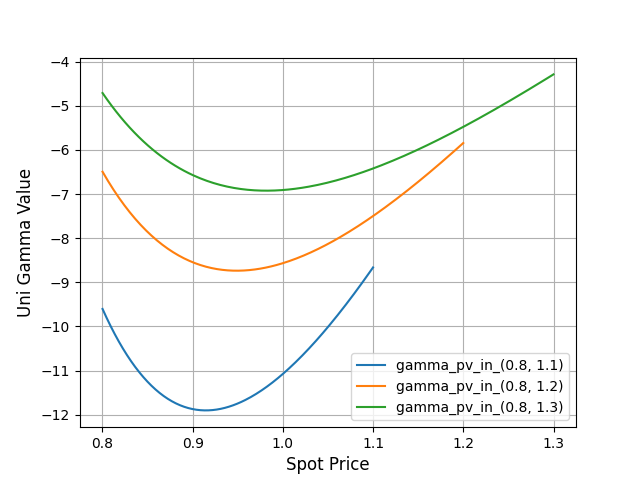}
    \caption{Different Range LP Pricing gamma Comparison}
    \label{fig:Different Range LP Pricing gamma Comparison}
\end{figure}

\begin{figure}
    \centering
    \includegraphics[width=1\linewidth]{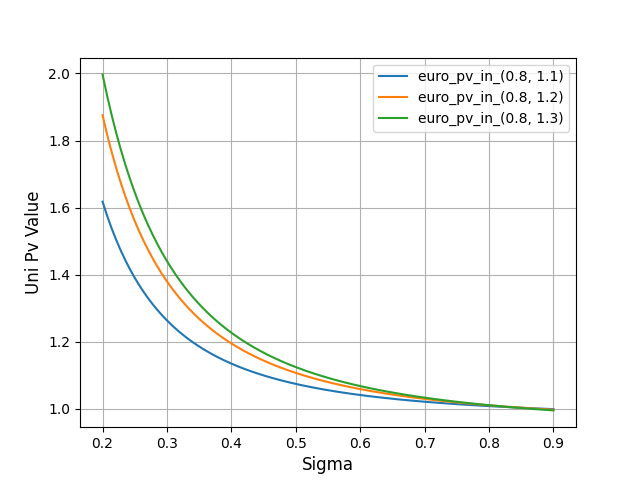}
    \caption{Different Range LP Pricing Pricing Analyiss versus vol}
    \label{fig:Different Range LP Pricing Pricing Analyiss versus vol}
\end{figure}

From Figure~\ref{fig:Different Range LP Pricing Comparison}, wider ranges increase European LP valuations, as assets stay longer within the range, extending fee collection time. Figure~\ref{fig:Different Range LP Pricing delta Comparison} shows delta rises with range width, as price fluctuations impact exchanges less relative to total position value, strengthening price correlation. Conversely, narrow ranges weaken this correlation, reducing delta.

Figure~\ref{fig:Different Range LP Pricing gamma Comparison} indicates that while wider ranges increase delta risk, gamma decreases, making delta hedging easier due to smoother delta changes. Finally, Figure~\ref{fig:Different Range LP Pricing Pricing Analyiss versus vol} confirms vega remains negative across ranges, highlighting the negative price-volatility relationship.

\subsection{The Impact of Volatility}

The relationship between underlying asset volatility and Uniswap V3 pricing is key to our sensitivity analysis. Intuitively, the European option model suggests PV should negatively correlate with volatility, as Uniswap V3's AMM acts as a short volatility strategy. Higher volatility thus lowers valuation.

\begin{figure}
    \centering
    \includegraphics[width=1\linewidth]{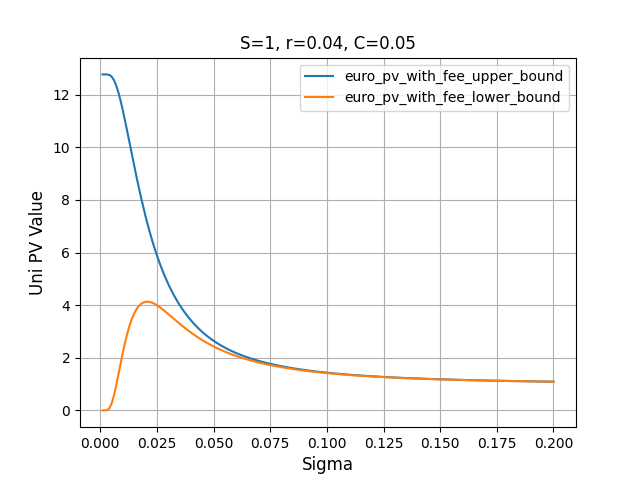}
    \caption{relation between pv and $\sigma$ using different rebate calculation}
    \label{fig:rebate_comparison}
\end{figure}

However, Figure~\ref{fig:rebate_comparison}'s orange line deviates due to the dual pricing model for fees. The lower bound reflects fee collection time, proportional to volatility via stopping time theory. The upper bound assumes continuous trading, justifying positive valuation in low volatility scenarios.

\begin{figure}
    \centering
    \includegraphics[width=1\linewidth]{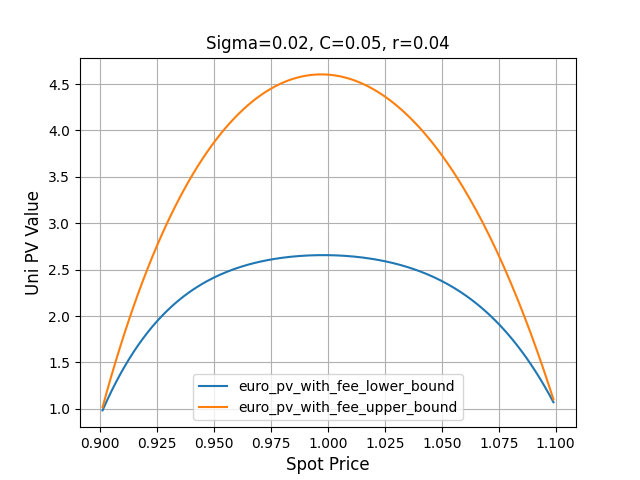}
    \caption{relationship of pv-spot with different fee calculation}
    \label{fig:different_fee_pv}
\end{figure}

Figure~\ref{fig:different_fee_pv} compares PV for $\sigma=0.02$. The upper-bound fee model consistently yields higher PV than the lower-bound model.

\section{Application of model}
In this section, we apply model to calculate the implied volatility and try to generalize the model to more specific AMM case.

\subsection{Implied Volatility}
Building upon the foundation of pricing research, we can also deduce the implied volatility of corresponding Uni V3 positions by inversely solving the pricing model.We can aggregate the implied volatility views of all Uni V3 investors at a cross-section, weighting them to observe the historical performance of the weighted implied volatility views. Notably, when we compare the historical performance of implied volatility with LVR model, we uncover some fascinating patterns. 

The volatility inferred by LVR is based on the assumption of no arbitrage. We possess an instantaneous LVR formula, and we also understand that the collection of fees is approximately a linear model over an extremely short period. According to the \cite{b14}, under the no-arbitrage assumption, the instantaneous LVR should equate to the instantaneous fees collected.

$$
\begin{aligned}
C\lambda=\frac{\lambda \sigma^2}{4}
\end{aligned}
$$

Based on this established equality, the sigma term within the instantaneous LVR formula can be inversely calculated by other parameters in the equation, thereby yielding an implied volatility formula based on LVR.

$$
\begin{aligned}
\sigma=2\sqrt{C}
\end{aligned}
$$

\begin{figure}
    \centering
    \includegraphics[width=1\linewidth]{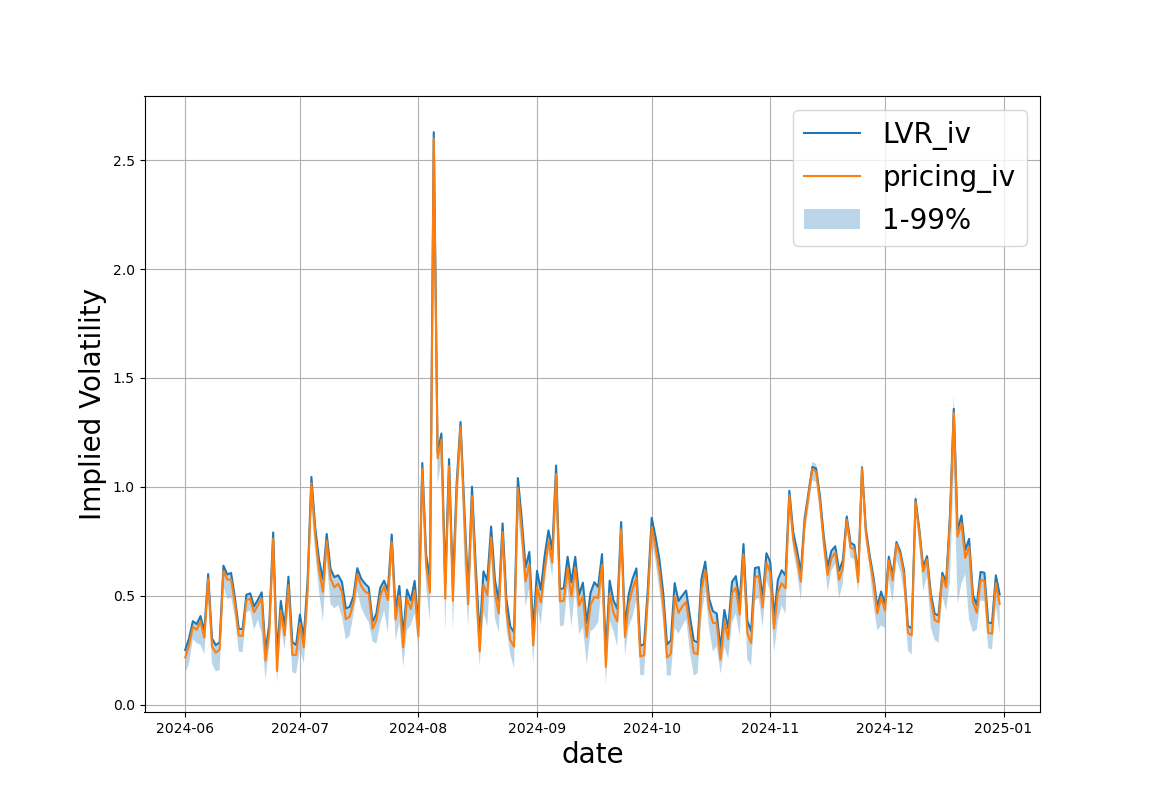}
    \caption{IV histogram Comparison}
    \label{fig:different_IV_histogram}
\end{figure}

 We collected historical transaction data for the ETH-USDC-005 trading pair on the Ethereum blockchain from June to December 2024 and created a comparative chart of implied volatility, the figure~\ref{fig:different_IV_histogram}. We can observe that the historical performance of the weighted IV largely aligns with the trend of the IV inferred by LVR. Additionally, the graph presents unweighted and non-aggregated IV samples, revealing that the IV deduced by LVR predominantly resides at the upper echelon of the IV samples. This phenomenon can be attributed to our model's assumption that future returns are subject to discounting. Consequently, our model inherently presumes the advantage of early boundary attainment for profit realization. As a result, the IV derived from our model is comparatively underestimated relative to the IV inferred by LVR.

While the implied volatility expressions are notably similar across the three models, key differences lie in their assumptions and specifics:

\begin{enumerate}
    \item Lambert's model assumes theta approximates a Dirac delta as time approaches zero, deriving the final option price. LVR also relies on infinitesimal time. Our model uses stopping time of a stochastic process to determine expiration, providing expected maturity time.
    
    \item LVR and Lambert assume zero risk-free rate, equating future earnings to present value. Our model incorporates the risk-free rate for more accurate present value estimation of Uni V3 positions.
    
    \item LVR and Lambert can only give the instantaneous value of Uni V3 position, in hence they can't capture greeks. On the contrary, our model can easily calculate greeks to measure the risk.
    
    \item LVR and Lambert focus on liquidity pool pricing, ignoring range order differences. We argue range orders reflect investor views on price and volatility, making price range relevant to Uni V3 pricing
    
    \item LVR and Lambert assume zero-drift geometric Brownian motion. Our risk-neutral model allows flexible drift assumptions.

    \item Like LVR, our model excels in scalability, as demonstrated in the next subsection.
\end{enumerate}

We summarize three models as following table, RNP represent our model's name risk-neutral-pricing model.

\begin{flushleft}
    \begin{tabular}{|c|c|c|c|}
    \hline
     & RNP  & Lambert  & LVR   \\
    \hline
    Maturity & stopping time   & 0   & flexible \\
    \hline
    risk free & flexible   & 0   & 0 \\
    \hline
    greeks & support & non-support & non-support \\
    \hline
    AMM range & relevant & irrelevant & irrelevant \\
    \hline
    price process & GBM & 0 drift GBM & 0 drift GBM \\
    \hline
    scalability & other AMM & none & other AMM \\
    \hline
    \end{tabular}
\end{flushleft}

\subsection{Extend to other AMM case}
\subsubsection{Example of application in Uniswap V2}
In the case of Uniswap V2, as the liquidity bounds (LH) approach zero and infinity, the stopping time for the euro-situation tends toward infinity, rendering the model ineffective. However, the modeling approach for the amer-situation can still be applied to Uniswap V2. This is because, as the duration of the position extends, the discounted value of the collected fees has a diminishing impact on overall returns, making fee collection less attractive to investors. The price range at which investors choose to exit their Uniswap V2 positions influences the expected duration of the position. Therefore, based on the price range at which investors actively close their positions, our model can still price the Uniswap V2 position. 

As the formula represent:
$$
\begin{aligned}
& L_1<x<L_2 \\
& d = \frac{\ln L_2}{\sigma}, d' = d-x\\
&c = \frac{\ln L_1}{\sigma}, c' = x-c\\
\end{aligned}
$$
So we have the formula:
$$
\begin{aligned}
V_E &= LP_{L_2}^{V2\ case} \exp({\mu'd'})\frac{\sinh[c'\sqrt{\mu'^2+2r}]}{\sinh[(d-c)\sqrt{\mu'^2+2r}]}\\
&+LP_{L_1}^{V2\ case}\exp({-\mu'c'})\frac{\sinh[(d'\sqrt{\mu'^2+2r}]}{\sinh[(d-c)\sqrt{\mu'^2+2r}]}    
\end{aligned} 
$$
This formula can solve the Uniswap V2 pricing problem.

\subsubsection{Application in Uniswap V4: Dynamic fee example}
Uniswap V4 has a unique mechanism called dynamic fee. Unlike V3's constant fee. Dynamic fee can be setting more flexible. One kind of dynamic fee can be adjusted in real-time based on various market conditions, like volatility. It means we can regard fee as a variable that is directly influenced by volatility. Our model can easily adapt dynamic fee mechanism, by adjusting the internal parameter C of the model into a function of volatility. We believe our model will also be highly adaptable to the pricing of Uniswap V4 positions for variety hooks.

\section{Conclusion}
In this paper, we have constructed a risk-neutral pricing model based on the stopping time theorem to price Uniswap V3 positions. Our model capture the value of each investor's range order is predicated on the discount of all potential earnings during the period until they first exit the position. From this hypothesis, we regard the Uniswap V3 position as an option pricing problem and have derived several analytical solution based on euro-situation and amer-situation, along with the corresponding Greeks based on these pricing formulas. The Greeks can be utilized to measure the risk exposure faced by investors holding this position. We have also employed MC simulations to verify the accuracy of our formula. This pricing model successfully delineates the relationship between the value of a Uniswap V3 position, its range setting, and the market's volatility. Through this model, investors can discern the implied volatility perspective underlying their Uni V3 positions and compare it with the prevailing liquidity views in the market. Moreover, with an accurate estimation of risk exposure, investors can more readily manage their risks and steer clear of undesired exposures.

\vspace{12pt}

\end{document}